\documentclass[11pt,fleqn,nofootinbib,superscriptaddress,floatfix]{revtex4}
\usepackage[latin9]{inputenc}
\usepackage{color}
\usepackage{amsmath}
\usepackage{amssymb}
\usepackage{graphicx}
\usepackage[unicode=true,
 bookmarks=false,
 breaklinks=false,pdfborder={0 0 1},backref=section,colorlinks=false]
 {hyperref}
\hypersetup{
 colorlinks,bookmarksopen,bookmarksnumbered,linkcolor=blu,pdfstartview=FitH,urlcolor=rossos,citecolor=rossos}

\usepackage[compat=1.1.0]{tikz-feynman}
\usepackage{textcomp}\usepackage{amsfonts}\usepackage{graphics}\usepackage{epstopdf}\usepackage{slashed}\usepackage{multirow}\usepackage{adjustbox}\usepackage{lipsum}\usepackage{rotating}\usepackage{xcolor}\usepackage{wrapfig}\usepackage{epsfig}
\usepackage[justification=raggedright, margin=5mm,font=sf,small,labelfont=bf, format=plain]{caption}
\definecolor{rosso}{cmyk}{0,1,1,0.4}
\definecolor{rossos}{cmyk}{0,1,1,0.55}
\definecolor{rossoc}{cmyk}{0,1,1,0.2}
\definecolor{blu}{cmyk}{1,1,0,0.3}
\definecolor{blus}{cmyk}{1,1,0,0.6}
\definecolor{bluc}{cmyk}{1,1,0,0.1}
\definecolor{verde}{cmyk}{0.92,0,0.59,0.25}
\definecolor{verdec}{cmyk}{0.92,0,0.59,0.15}
\definecolor{verdes}{cmyk}{0.92,0,0.59,0.4}
\definecolor{lime}{HTML}{A6CE39}
\DeclareRobustCommand{\orcidicon}{\hspace{-2.1mm}
\begin{tikzpicture}
\draw[lime,fill=lime] (0,0.0) circle [radius=0.13] node[white] {{\fontfamily{qag}\selectfont \tiny ID}}; \draw[white,fill=white] (-0.0525,0.095) circle [radius=0.007];
\end{tikzpicture} \hspace{-3.7mm} }
\foreach \x in {A, ..., Z}
 {\expandafter\xdef\csname orcid\x\endcsname{\noexpand\href{https://orcid.org/\csname orcidauthor\x\endcsname} {\noexpand\orcidicon}}}

\begin{document}
\title{\color{bluc}Constraining the Georgi-Machacek Model with a Light Higgs}
\author{Amine Ahriche\orcidA{}}
\email{ahriche@sharjah.ac.ae}

\affiliation{Department of Applied Physics and Astronomy, University of Sharjah,
P.O. Box 27272 Sharjah, UAE.}
\begin{abstract}
In this work, we investigate the viability of a light Higgs ($\eta$)
scenario in the Georgi-Machacek (GM) model, where we consider all
theoretical and experimental constraints such as the perturbativity,
vacuum stability, unitarity, electroweak precision tests, the Higgs
di-photon and undetermined decays and the Higgs total decay width.
In addition, we consider more recent experimental bounds from the
searches for doubly-charged Higgs bosons in the VBF channel $H_{5}^{++}\to  W^{+}W^{+}$,
Drell-Yan production of a neutral Higgs boson $pp\to  H_{5}^{0}(\gamma\gamma)H_{5}^{+}$,
and for the light scalars at LEP $e^{-}e^{+}\to  Z\eta$, and
at ATLAS and CMS in different final states such as $pp\to \eta\to 2\gamma$
and $pp\to  h\to \eta\eta\to 4\gamma,2\mu2\tau,2\mu2b,2\tau2b$.
By combining these bounds together, we found a parameter space region
that is significant as the case of the SM-like Higgs to be the light
CP-even eigenstate, and this part of the parameter space would be
tightened by the coming analyses.
\end{abstract}
\maketitle

\section{Introduction}

Despite the discovery of a Standard Model (SM)-like 125 GeV Higgs
boson at the Large Hadron Collider (LHC)~\cite{ATLAS:2012yve,CMS:2012qbp},
the SM is not complete since it can not provide answers to many questions,
such as the hierarchy problem, fermions masses, strong CP problem,
the dark matter nature, baryon asymmetry at the Universe; and the
neutrino oscillation data. The discovered 125 GeV scalar seems to
be the SM Higgs, however, it is not clear how does the electroweak
symmetry breaking (EWSB) occurs: via one single scalar field or more?

In many SM extensions, the EWSB is achieved by more than one scalar,
where some of the new scalar fields acquire vacuum expectation values
(VEV), and then mix with the SM doublet, which make the 125 GeV Higgs
is a composite field. Among the popular models, the so-called Georgi-Machacek
(GM) model~\cite{Georgi:1985nv}, where the SM is extended by one
complex and one real scalar triplets, assigned by a global custodial
$SU(2)_{V}$ symmetry, that is preserved in the scalar potential after
the EWSB. The model vacuum is defined in a way that predicts a tree-level
custodial symmetry, under which the scalar spectrum manifests in multiplets:
a quintet ($H_{5}$), a triplet ($H_{3}$) and two CP-even singlets
($\eta$ and $h$).

The existence of extra scalar degrees of freedom and the non-trivial
couplings of the Higgs to the gauge bosons, make GM model phenomenologically
very rich~\cite{Chanowitz:1985ug,Gunion:1989ci,Haber:1999zh,Aoki:2007ah,Godfrey:2010qb,Low:2010jp,Logan:2010en,Chang:2012gn,Kanemura:2013mc,Englert:2013zpa,Killick:2013mya,Englert:2013wga,Ghosh:2019qie,Das:2018vkv,Hartling:2014zca,Hartling:2014aga,Chiang:2014bia,Chiang:2015rva,Chang:2017niy,Chiang:2015kka,Chen:2022zsh}.
In addition, the GM model can address some the SM open questions,
such as the neutrino mass~\cite{Chen:2020ark}, dark matter~\cite{Pilkington:2017qam},
and the electroweak phase transition strength~\cite{Chiang:2014hia}.
The GM scalar sector has been confronted with the existing data~\cite{Ismail:2020zoz},
where direct search constraints for extra Higgs bosons and measurements
of the SM-like Higgs properties are considered. In addition, the authors
derived bounds from the negative searches of the doubly-charged Higgs
bosons in the VBF channel $H_{5}^{++}\to  W^{+}W^{+}$; and
the Drell-Yan production of a neutral Higgs boson $pp\to  H_{5}^{0}(\gamma\gamma)H_{5}^{+}$.
In~\cite{Bairi:2022adc}, we have investigated the GM parameter space
where the SM-like Higgs is considered to be the light CP-even eigenstate
$h=h_{125}$; and the eigenstate $\eta$ is a heavier scalar $m_{\eta}>m_{h}$.
We have considered all the known theoretical and experimental constraints,
including those from the negative searches of the heavy scalar via
$pp\to \eta\to  hh,\tau\tau,ZZ$; and we have found
that a significant part of the parameter space is viable and could
be probed soon by future analyses with more data. One has to mention
that it turns out that two thirds of the parameter space allowed by
the constraints described in the literature, are excluded by some
possibly existing scalar potential minima (that either preserve or
violate the CP and electric charge symmetries) that are deeper than
the electroweak (EW) vacuum~\cite{Bairi:2022adc}. Here, we aim to
investigate other part of the parameter space where the SM-like Higgs
is the heavy CP-even eigenstate, i.e., $h=h_{125}$ and $m_{\eta}<m_{h}$.
We will consider all the above mentioned constraints in addition to
constraints from the searches of light CP-even scalar whether are
direct at LEP $e^{-}e^{+}\to  Z\eta$, and LHC $pp\to \eta\to \gamma\gamma$~\cite{CMS:2018cyk,ATLAS:2018xad,ATLAS:2022nkn};
or indirect $pp\to  h\to \eta\eta\to  X\overline{X}Y\overline{Y}$~\cite{CMS:2018zvv,CMS:2018qvj,CMS:2022xxa,ATLAS:2021ypo,ATLAS:2021hbr}.

In section~\ref{sec:Model}, we review the GM model and present the
constraints described in the literature. We discuss the physics of
a light CP-even scalar at colliders in section~\ref{sec:Eta}; and
discuss our numerical results in section~\ref{sec:NA}. In section~\ref{sec:Conclusion},
we give our conclusion.

\section{The Model: Parameters and Constraints\label{sec:Model}}

The GM model scalar sector includes a doublet $(\phi^{+},\,\phi^{0})^{T}$;
and two triplets $(\chi^{++},\,\chi^{+},\,\chi^{0})^{T}$ and $(\xi^{+},\,\xi^{0},\,-\xi^{-})^{T}$
with the hypercharge $Y=1,2,0$, respectively;
\begin{equation}
\Phi=\left(\begin{array}{cc}
\frac{\upsilon_{\phi}+h_{\phi}-ia_{\phi}}{\sqrt{2}} & \phi^{+}\\
-\phi^{-} & \frac{\upsilon_{\phi}+h_{\phi}+ia_{\phi}}{\sqrt{2}}
\end{array}\right),\quad\Delta=\left(\begin{array}{ccc}
\frac{\upsilon_{\chi}+h_{\chi}-ia_{\chi}}{\sqrt{2}} & \xi^{+} & \chi^{++}\\
-\chi^{-} & \upsilon_{\xi}+h_{\xi} & \chi^{+}\\
\chi^{--} & -\xi^{-} & \frac{\upsilon_{\chi}+h_{\chi}+ia_{\chi}}{\sqrt{2}}
\end{array}\right),\label{eq:field}
\end{equation}
where the vacuum expectation values (VEV's) satisfy the relations
$\upsilon_{\chi}=\sqrt{2}\upsilon_{\xi}$ and $\upsilon_{\phi}^{2}+8\upsilon_{\xi}^{2}\equiv\upsilon^{2}=(246.22\,\mathrm{GeV})^{2}$;
to ensure the tree-level custodial symmetry. The scalar potential
that is invariant under the global symmetry $SU(2)_{L}\times SU(2)_{R}\times U(1)_{Y}$
in the GM model is
\begin{align}
V(\varPhi,\Delta) & =\frac{m_{1}^{2}}{2}\mathrm{Tr}[\varPhi^{\dagger}\varPhi]+\frac{m_{2}^{2}}{2}\mathrm{Tr}[\Delta^{\dagger}\Delta]+\lambda_{1}(\mathrm{Tr}[\varPhi^{\dagger}\varPhi])^{2}+\lambda_{2}\mathrm{Tr}[\varPhi^{\dagger}\varPhi]\mathrm{Tr}[\Delta^{\dagger}\Delta]+\lambda_{3}\mathrm{Tr}[(\Delta^{\dagger}\Delta)^{2}]\nonumber \\
 & +\lambda_{4}(\mathrm{Tr}[\Delta^{\dagger}\Delta])^{2}-\lambda_{5}\mathrm{Tr}[\varPhi^{\dagger}\frac{\sigma^{a}}{2}\varPhi\frac{\sigma^{b}}{2}]\mathrm{Tr}[\Delta^{\dagger}T^{a}\Delta T^{b}]-\mu_{1}\mathrm{Tr}[\varPhi^{\dagger}\frac{\sigma^{a}}{2}\varPhi\frac{\sigma^{b}}{2}](U\Delta U^{\dagger})_{ab}\nonumber \\
 & -\mu_{2}\mathrm{Tr}[\Delta^{\dagger}T^{a}\Delta T^{b}](U\Delta U^{\dagger})_{ab},\label{eq:V}
\end{align}
with $\sigma^{1,2,3}$ are the Pauli matrices and $T^{1,2,3}$ correspond
to the generators of the $SU(2)$ triplet representation and the matrix
$U$ is given in~\cite{Georgi:1985nv}. After the EWSB, we are left
with: three CP-even eiegenstaes $\{h,\eta,H_{5}^{0}\}$, one CP-odd
eigenstate $H_{3}^{0}$, two singly charged scalars $\{H_{3}^{\pm},H_{5}^{\pm}\}$,
and one doubly charged scalar $H_{5}^{\pm\pm}$, that are defined
as
\begin{align}
h & =c_{\alpha}h_{\phi}-\frac{s_{\alpha}}{\sqrt{3}}(\sqrt{2}h_{\chi}+h_{\xi}),\,\eta=s_{\alpha}h_{\phi}+\frac{c_{\alpha}}{\sqrt{3}}(\sqrt{2}h_{\chi}+h_{\xi}),\,H_{5}^{0}=\sqrt{\frac{2}{3}}h_{\xi}-\sqrt{\frac{1}{3}}h_{\chi},\nonumber \\
H_{3}^{0} & =-s_{\beta}a_{\phi}+c_{\beta}a_{\chi},\,H_{3}^{\pm}=-s_{\beta}\phi^{\pm}+c_{\beta}\frac{1}{\sqrt{2}}(\chi^{\pm}+\xi^{\pm}),\,H_{5}^{\pm}=\frac{1}{\sqrt{2}}(\chi^{\pm}-\xi^{\pm}),\,H_{5}^{\pm\pm}=\chi^{\pm\pm},\label{eq:Eigen}
\end{align}
with $\tan\beta=2\sqrt{2}\upsilon_{\xi}/\upsilon_{\phi}$ and $\tan2\alpha=2M_{12}^{2}/(M_{22}^{2}-M_{11}^{2})$,
where $M^{2}$ is the mass squared matrix in the basis $\{h_{\phi},\,\sqrt{\frac{2}{3}}h_{\chi}+\frac{1}{\sqrt{3}}h_{\xi}\}$.
The quartic couplings $\lambda$'s can be expressed as
\begin{align}
\lambda_{1} & =\frac{\varrho_{1}c_{\alpha}^{2}+\varrho_{2}s_{\alpha}^{2}}{8\upsilon^{2}c_{\beta}^{2}},\,\lambda_{2}=\frac{m_{3}^{2}}{\upsilon^{2}}-\frac{c_{\alpha}s_{\alpha}}{\sqrt{6}\upsilon^{2}c_{\beta}s_{\beta}}(\varrho_{1}-\varrho_{2})-\frac{\mu_{1}}{2\sqrt{2}\upsilon s_{\beta}},\nonumber \\
\lambda_{3} & =\frac{\sqrt{2}(\mu_{1}c_{\beta}^{2}-3\mu_{2}s_{\beta}^{2})}{s_{\beta}^{3}\upsilon}-\frac{3c_{\beta}^{2}m_{3}^{2}}{s_{\beta}^{2}\upsilon^{2}}+\frac{m_{5}^{2}}{s_{\beta}^{2}\upsilon^{2}},\,\lambda_{5}=-\frac{\sqrt{2}\mu_{1}}{\upsilon s_{\beta}}+\frac{2m_{3}^{2}}{\upsilon^{2}},\nonumber \\
\lambda_{4} & =-\frac{\mu_{1}c_{\beta}^{2}-3\mu_{2}s_{\beta}^{2}}{\sqrt{2}s_{\beta}^{3}\upsilon}+\frac{c_{\beta}^{2}m_{3}^{2}}{s_{\beta}^{2}\upsilon^{2}}-\frac{m_{5}^{2}}{3s_{\beta}^{2}\upsilon^{2}}+\frac{\varrho_{1}s_{\alpha}^{2}+\varrho_{2}c_{\alpha}^{2}}{3s_{\beta}^{2}\upsilon^{2}},\label{eq:lm}
\end{align}
with $\varrho_{1}=\min(m_{h}^{2},m_{\eta}^{2})$ and $\varrho_{2}=\max(m_{h}^{2},m_{\eta}^{2})$.
The formulas of $\lambda_{1,2,4}$ here are valid for both cases of
$m_{h}<m_{\eta}$ and $m_{h}>m_{\eta}$.

It has been shown in~\cite{Bairi:2022adc} that the scalar potential
(\ref{eq:V}) could acquire some minima that could violate the CP-symmetry
and/or electric charge, where they could be deeper than the electroweak
vacuum $\{\upsilon_{\phi},\sqrt{2}\upsilon_{\xi},\upsilon_{\xi}\}$.
Then, this part of the parameter space would be ignored. Here, we
impose the constraints from (1) vacuum stability, (2) unitarity, (3)
the electroweak precision tests, (4) the di-photon and undetermined
Higgs branching ratios and total decay width; in addition to (5) the
constraints from negative searches for light scalar resonances at
LEP~\cite{OPAL:2002ifx}. For the constraints (1-3), we used the
results described in~\cite{Bairi:2022adc}.

In this setup, the SM-like Higgs $h$ (the CP-even scalar $m_{h}=125.18\,\mathrm{GeV}$)
decays mainly into pairs of fermions $(cc,\,\mu\mu,\,\tau\tau,\,b\overline{b})$
and gauge bosons $WW^{*}$ and $ZZ^{*}$, in addition to a pair of
light scalars $\eta\eta$ when kinematically allowed. Since the Higgs
couplings to SM fields are scaled by the coefficients
\begin{align}
\kappa_{\mathfrak{\mathrm{F}}} & =\frac{g_{hff}^{GM}}{g_{hff}^{SM}}=\frac{c_{\alpha}}{c_{\beta}},\,\kappa_{V}=\frac{g_{hVV}^{GM}}{g_{hVV}^{SM}}=c_{\alpha}c_{\beta}-\sqrt{\frac{8}{3}}s_{\alpha}s_{\beta},\label{eq:K}
\end{align}
then, its total decay width can be written as
\begin{align}
\varGamma_{h}^{tot} & =\Gamma_{h}^{SM}\sum_{X=SM}\kappa_{X}^{2}\mathcal{B}^{SM}(h\to  XX)+\Theta(m_{h}-2m_{\eta})\frac{g_{h\eta\eta}^{GM}}{32\pi m_{h}}\big(1-4m_{\eta}^{2}/m_{h}^{2}\big)^{1/2},\label{eq:GM-Gamma}
\end{align}
where the last term represents the partial decay width $\Gamma(h\to \eta\eta)$,
$\varGamma_{h}^{SM}=4.08\,\mathrm{MeV}$~\cite{ParticleDataGroup:2020ssz}
and $\mathcal{B}^{SM}(h\to  XX)$ are the SM values for total
decay width and the branching ratios for the Higgs, respectively.
Here, $g_{h\eta\eta}^{GM}$ is the scalar triple coupling $h\eta\eta$.
Since the light scalar $\eta$ can be seen at detectors via its decay
to light fermions $\eta\to  f\bar{f}$, then, the Higgs decay
$h\to \eta\eta$ does not match any of the known SM final states,
and hence called undetermined channel, which is constrained by ATLAS
as $\mathcal{B}_{und}<0.22$~\cite{ATLAS:2019cid,ATLAS:2019nkf}.
The total Higgs decay width recent measurements based on the off-shell
Higgs production in the final state $h\to  ZZ^{*}\to \ell\ell\nu\nu$
give the upper bound $\Gamma_{h}=4.6_{-2.5}^{+2.6}~\textrm{MeV}$~\cite{ATLAS-CONF-2022-068}.

In previous analysis of the GM parameter space~\cite{Bairi:2022adc},
it has been shown that the measurements of the Higgs signal strength
modifiers imply constraints on the coefficients $\kappa_{F,V}$. Here,
another factor is constrained in addition, which is the undetermined
Higgs decay; and consequently the scalar $\eta$ mass and the triple
coupling $g_{h\eta\eta}^{GM}$. Then, the partial Higgs signal strength
modifier for the channel $h\to XX$ ($X=f,V$) can be simplified in
this setup as
\begin{equation}
\mu_{XX}=\frac{\sigma(pp\to  h)\times\mathcal{B}(h\to  XX)}{\sigma^{SM}(pp\to  h)\times\mathcal{B}^{SM}(h\to  XX)}=\kappa_{F}^{2}\kappa_{X}^{2}(1-\mathcal{B}_{und}),\label{eq:muXX}
\end{equation}
where $\sigma(gg\to  h)\,[\sigma^{SM}(gg\to  h)]$ is
the $ggF$ production cross section in the GM [SM] model.

For the di-photon Higgs strength modifier, it can be written as
\begin{align}
\mu_{\gamma\gamma}=\kappa_{F}^{2}(1-\mathcal{B}_{und})\left|\frac{\frac{\upsilon}{2}\sum_{i}\frac{g_{hii}^{GM}}{m_{i}^{2}}Q_{i}^{2}A_{0}^{\gamma\gamma}(4m_{i}^{2}/m_{h}^{2})+\kappa_{V}A_{1}^{\gamma\gamma}(4m_{W}^{2}/m_{h}^{2})+\kappa_{F}\frac{4}{3}A_{1/2}^{\gamma\gamma}(4m_{t}^{2}/m_{h}^{2})}{A_{1}^{\gamma\gamma}(4m_{W}^{2}/m_{h}^{2})+\frac{4}{3}A_{1/2}^{\gamma\gamma}(4m_{t}^{2}/m_{h}^{2})}\right|^{2},\label{eq:Rgg}
\end{align}
where $i=H_{3}^{+},\,H_{5}^{+},\,H_{5}^{++}$ stands for all charged
scalars inside the loop diagrams, $Q_{i}$ is the electric charge
of the field $i$ in units of $|e|$, $g_{hii}^{GM}$ are the Higgs
triple couplings to the charged scalars; and the functions $A_{i}^{\gamma\gamma}$
are given in the literature~\cite{Djouadi:2005gi}. According to
the partial Higgs strength modifier formulas in (\ref{eq:muXX}) and
(\ref{eq:Rgg}), one expects that the experimental constraints would
enforce the coefficients $\kappa_{F,V}$ to lie around unity. In our
numerical scan, we consider the recent experimental values~\cite{ParticleDataGroup:2022pth}
for the partial Higgs strength modifiers (\ref{eq:muXX}) and (\ref{eq:Rgg}).

Besides the above mentioned constraints, the negative searches for
doubly-charged Higgs bosons in the VBF channel $H_{5}^{++}\to  W^{+}W^{+}$;
and from Drell-Yan production of a neutral Higgs boson $pp\to  H_{5}^{0}(\gamma\gamma)H_{5}^{+}$;
give strong bounds on the parameter space~\cite{Ismail:2020zoz}.
It has been shown in~\cite{Ismail:2020zoz}, that the doubly-charged
Higgs bosons in the VBF channel leads to a constraint from CMS on
$s_{\beta}^{2}\times\mathcal{B}(H_{5}^{++}\to  W^{+}W^{+})$~~\cite{CMS:2017fhs}.
Clearly, non vanishing values for the branching ratios $\mathcal{B}(H_{5}^{++}\to  W^{+}H_{3}^{+})$
(for $m_{5}>m_{3}+m_{W}$) and $\mathcal{B}(H_{5}^{++}\to  H_{3}^{+}H_{3}^{+})$
(for $m_{5}>2m_{3}$), will significantly help to relax this bounds.
While the relevant quantity for the constraints on $H_{5}^{0}\to \gamma\gamma$
is the fiducial cross section times branching ratio $\sigma_{fid}=\big(\sigma_{H_{5}^{0}H_{5}^{+}}\times\epsilon_{+}+\sigma_{H_{5}^{0}H_{5}^{-}}\times\epsilon_{-}\big)\mathcal{B}(H_{5}^{0}\to \gamma\gamma)$,
that is constrained by ATLAS at 8 TeV~\cite{ATLAS:2014jdv} and at
13 TeV~\cite{ATLAS:2017ayi}. Here, we used the decay rate formulas,
the cross section and efficiency values used in~\cite{Ismail:2020zoz}
to include these constraints in our numerical analysis.

Since part of the charged triplet $H_{3}^{\pm}$ is coming from the
SM doublet as shown in (\ref{eq:Eigen}), then it should couple the
up and down quarks as the W gauge bosons does. This interaction leads
to flavor violating processes such as the $b\to s$ transition ones,
which depend only on the charged triplet mass $m_{3}$ and the mixing
angle $\beta$ (and consequently the triplet VEV $\upsilon_{\chi}$).
In our numerical scan, we consider the bounds on the $m_{3}$-$\upsilon_{\chi}$
plan shown in~\cite{Hartling:2014aga}.

\section{the light scalar $\eta$ in the Collider\label{sec:Eta}}

After the discovery of the Higgs boson with $m_{h}=125.18\,\mathrm{GeV}$,
efforts have been devoted to search for light neutral scalar boson
through different channels over a wide range of mass. Such results
can also be used to impose constraints on models with many neutral
scalars such as the GM model.

The two CP-even eigenstates $h$ and $\eta$ are defined through a
mixing angle $\alpha$ and $(m_{\eta}<m_{h})$, where the heavy eigenstate
$h$ is identified to be the SM-like Higgs with the measured mass
$m_{h}=125.18\,\mathrm{GeV}$. Here, the light scalar $\eta$ has
similar couplings as the SM Higgs, but modified with the factors
\begin{align}
\zeta_{V} & =\frac{g_{\eta VV}^{GM}}{g_{hVV}^{SM}}=s_{\alpha}c_{\beta}+\sqrt{\frac{8}{3}}c_{\alpha}s_{\beta},\,\zeta_{F}=\frac{g_{\eta FF}^{GM}}{g_{hFF}^{SM}}=\frac{s_{\alpha}}{c_{\beta}},\label{eq:Zeta}
\end{align}
then, the partial decay width of the light scalar $\eta$ into SM
final states can be written as $\Gamma(\eta\to  X\overline{X})=\zeta_{X}^{2}\Gamma^{SM}(\eta\to  X\overline{X})$,
where $\Gamma^{SM}(\eta\to  X\overline{X})$ is the Higgs partial
decay width estimated at $m_{h}\to  m_{\eta}$~\cite{Higgs}.
Thus, its total decay width can be written as
\begin{equation}
\varGamma_{\eta}^{tot}=\Gamma_{\eta}^{SM}\sum_{X=SM}\zeta_{X}^{2}\mathcal{B}^{SM}(\eta\to  X\overline{X}),\label{eq:eta-Gamma}
\end{equation}
where $\Gamma_{\eta}^{SM}$ and $\mathcal{B}^{SM}(\eta\to  X\overline{X})$
are the Higgs total decay width and the branching ratios estimated
at $m_{h}\to  m_{\eta}$~\cite{Higgs}. At colliders, there
have been many searches for light scalars which are translated into
constraints on the light scalar mass and its couplings to SM particles.
In what follows, we will focus on two types of searches of the light
scalar $\eta$: (1) direct production like $e^{-}e^{+},pp\to \eta+X$,
where the scalar could be identified via one of its SM-like decays
$\eta\to \gamma\gamma,\mu\mu,\tau\tau,cc,bb$, and (2) indirect
production via the Higgs decay $pp\to  h\to \eta\eta\to  X\overline{X}Y\overline{Y}$,
where the light scalar is identified via its SM decays $X,Y=\gamma,\mu,\tau,c,b$.
Here, we will consider the constraints from the negative searches
for $pp\to \eta\to \gamma\gamma$ at CMS at 8+13 TeV~\cite{CMS:2018cyk},
and at ALTAS at 13 TeV with integrated luminosity $80~\mathrm{fb}^{-1}$~\cite{ATLAS:2018xad};
and at $138~\mathrm{fb}^{-1}$~\cite{ATLAS:2022nkn}.

At LEP, many searches for Higgs at low mass range $m_{h}<100~\textrm{GeV}$
have been performed, and bounds on the form factor~\cite{OPAL:2002ifx};
that can be simplified in our setup as
\begin{equation}
\kappa_{Z\eta}=\frac{\sigma(e^{-}e^{+}\to ~\eta~Z)}{\sigma^{SM}(e^{-}e^{+}\to ~\eta~Z)}=\zeta_{V}^{2}.\label{eq:kappa}
\end{equation}

A similar search for light SM-like Higgs in the di-photon channel
with masses in the range $70-110~\mathrm{GeV}$ has been done by CMS
at 8 $\mathrm{TeV}$ and 13 $\mathrm{TeV}$~\cite{CMS:2018cyk},
where some upper bounds are established on the production cross section
scaled by its SM value $\zeta_{F}^{2}.\kappa_{\gamma\gamma}^{\eta}=\frac{\sigma(pp\to \eta)\times\mathcal{B}(\eta\to \gamma\gamma)}{\sigma^{SM}(pp\to \eta)\times\mathcal{B}^{SM}(\eta\to \gamma\gamma)}$.
In our setup, the parameter $\kappa_{\gamma\gamma}^{\eta}$ can be
simplified as
\begin{equation}
\kappa_{\gamma\gamma}^{\eta}=\zeta_{F}^{2}\frac{\Gamma_{SM}^{tot}(\eta)}{\Gamma^{tot}(\eta)}\left|\frac{\frac{\upsilon}{2}\sum_{i}\frac{g_{\eta ii}^{GM}}{m_{i}^{2}}Q_{i}^{2}A_{0}^{\gamma\gamma}\big(4m_{i}^{2}/m_{\eta}^{2}\big)+\zeta_{V}A_{1}^{\gamma\gamma}\big(4m_{W}^{2}/m_{\eta}^{2}\big)+\zeta_{F}\frac{4}{3}A_{1/2}^{\gamma\gamma}\big(4m_{t}^{2}/m_{\eta}^{2}\big)}{A_{1}^{\gamma\gamma}\big(4m_{W}^{2}/m_{\eta}^{2}\big)+\frac{4}{3}A_{1/2}^{\gamma\gamma}\big(4m_{t}^{2}/m_{\eta}^{2}\big)}\right|^{2},\label{eq:kgg}
\end{equation}
with $i=H_{3}^{\pm},~H_{5}^{\pm},~H_{5}^{\pm\pm}$, $g_{\eta ii}^{GM}$
are the scalar triple couplings of the scalar $\eta$ to the charged
scalars; and the functions $A_{0,1,1/2}^{\gamma\gamma}$ are given
in the literature~\cite{Djouadi:2005gi}.

Concerning the indirect searches via the Higgs decay into a pairs
of light scalars in the channels $h\to  HH\to  X\overline{X}Y\overline{Y}$
($H$ could be a CP-odd scalar $a$ or a CP-even one like $\eta$
in our setup), many searches by ATLAS and CMS at $8\,\mathrm{TeV}$
and/or $13\,\mathrm{TeV}$ and at different values for the integrated
luminosity values; have been performed in many final states, among
them that decay into photons or light fermions. For instance, negative
searches for $pp\to  h\to \eta\eta\to  X\overline{X}Y\overline{Y}$
established bounds either on the branching ratio $\mathcal{B}(h\to \eta\eta\to  X\overline{X}Y\overline{Y})$,
the ratio $\frac{\sigma(pp\to  h)}{\sigma^{SM}(pp\to  h)}\mathcal{B}(h\to \eta\eta\to  X\overline{X}Y\overline{Y})$
and/or the cross section section $\sigma(pp\to  h)\times\mathcal{B}(h\to \eta\eta\to  X\overline{X}Y\overline{Y})$.
In our work, we will consider the constraints coming from the CMS
and ATLAS searches for the Higgs decay into a pair of light pseudoscalars
in the final state $h\to 2a\to 2b2\tau$~\cite{CMS:2018zvv},
$h\to 2a\to 2\tau2\mu$ at $\sqrt{s}=13~\mathrm{TeV}$
and $\mathcal{L}=35.9~\mathrm{fb}^{-1}$~\cite{CMS:2018qvj}, $h\to 2a\to 4\gamma$
at CMS $\sqrt{s}=13~TeV$ and $\mathcal{L}=132~\mathrm{fb}^{-1}$~\cite{CMS:2022xxa},
$h\to 2a\to 2b2\mu$ at $\sqrt{s}=13~\mathrm{TeV}$
and $\mathcal{L}=36.1~\mathrm{fb}^{-1}$~\cite{ATLAS:2021ypo}, $h\to 2a\to 2b2\mu$
at ATLAS $\sqrt{s}=13~\mathrm{TeV}$ and $\mathcal{L}=139~\mathrm{fb}^{-1}$~\cite{ATLAS:2021hbr}.
All these analyses were performed using the ggF Higgs production mode
at the LHC.

\section{Numerical Analysis and Discussion\label{sec:NA}}

Here, we have considered the heavy CP-even scalar to be the $125\,\textrm{GeV}$
SM-like Higgs; and have taken into account the different theoretical
and experimental constraints described in Sections~\ref{sec:Model}
and~\ref{sec:Eta}, such as the constraints from perturbativity,
vacuum stability~\cite{Hartling:2014zca,Arhrib:2011uy}, electroweak
precision tests~\cite{Hartling:2014aga} , the di-photon and undetermined
Higgs decays, the total Higgs decay width; and the B physics flavor
constraints. In addition, we have considered also the constraints
from the fact that the EW vacuum $(\upsilon_{\phi},\,\sqrt{2}\upsilon_{\xi},\upsilon_{\xi})$
must be the deepest among possible minima that may preserve or violate
the CP and electric charge symmetries as described in~\cite{Bairi:2022adc}.
As a first step of this numerical study, we perform a full numerical
scan over the GM model parameter space, then, in a next step we will
imposes the constraints from the negative searches for doubly-charged
Higgs bosons in the VBF channel $H_{5}^{++}\to  W^{+}W^{+}$;
and from Drell-Yan production of a neutral Higgs boson $pp\to  H_{5}^{0}(\gamma\gamma)H_{5}^{+}$;
negative searches for a light Higgs at LEP $e^{-}e^{+}\to ~\eta~Z$~\cite{OPAL:2002ifx},
the direct search for a light resonance at the LHC~\cite{CMS:2018cyk}
and the indirect searches for light resonance via the final states
$h\to \eta\eta\to  X\overline{X}Y\overline{Y}$~\cite{CMS:2018zvv,CMS:2018qvj,CMS:2022xxa,ATLAS:2021ypo,ATLAS:2021hbr}.

The GM parameter space is described by the free parameters $\lambda_{2},\;\lambda_{4},\;m_{\eta},m_{3},\;m_{5},\,s_{\alpha}$
and $t_{\beta}=\tan\beta\equiv2\sqrt{2}\upsilon_{\xi}/\upsilon_{\phi}$,
within the ranges
\begin{equation}
78~\textrm{GeV}<m_{3,5}<3~\textrm{TeV},\,10~\textrm{GeV}<m_{\eta}<m_{h},\,\left|\lambda_{2,4}\right|\leq20,\,|t_{\beta}|\leq3,\label{eq:range}
\end{equation}
where for the lower bound on $m_{3,5}$, we considered lower mass
bound on this singly charged scalar obtained from the direct search
at the LEP II~\cite{LEPHiggsWorkingGroupforHiggsbosonsearches:2001ogs}.

Here, the negative values of $t_{\beta}$ are considered for the following
reason. In the GM model, there exists an invariance under the transformation
$(\upsilon_{\xi},\mu_{1,2})\to(-\upsilon_{\xi},-\mu_{1,2})$, which
means $V(\Phi,\Delta,\mu_{1,2})=V(\Phi,-\Delta,-\mu_{1,2})$. Consequently,
the scalar mass matrix elements also remain invariant under this transformation.
However, because the physical scalar eigenstates are mixtures of the
components of $\Phi$ and $\Delta$, most of the physical vertices
that involves scalars are not invariant under $(\upsilon_{\xi},\mu_{1,2})\to(-\upsilon_{\xi},-\mu_{1,2})$.
This means that these vertices change; and therefore any two benchmark
points (BPs) with the same input parameters but with different signs
of $(\pm t_{\beta},\pm\mu_{1,2})$ are physically different. This
can be seen in the scaling factors (\ref{eq:K}) and (\ref{eq:Zeta}).
This makes the BPs with negative $t_{\beta}$ values in (\ref{eq:range})
independent part of the parameter space that should not be ignored.

\begin{figure}[h]
\includegraphics[width=0.49\textwidth]{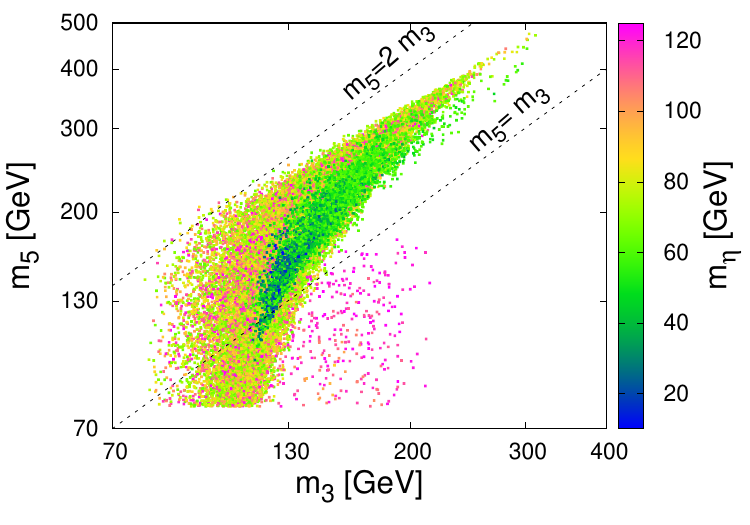}~\includegraphics[width=0.49\textwidth]{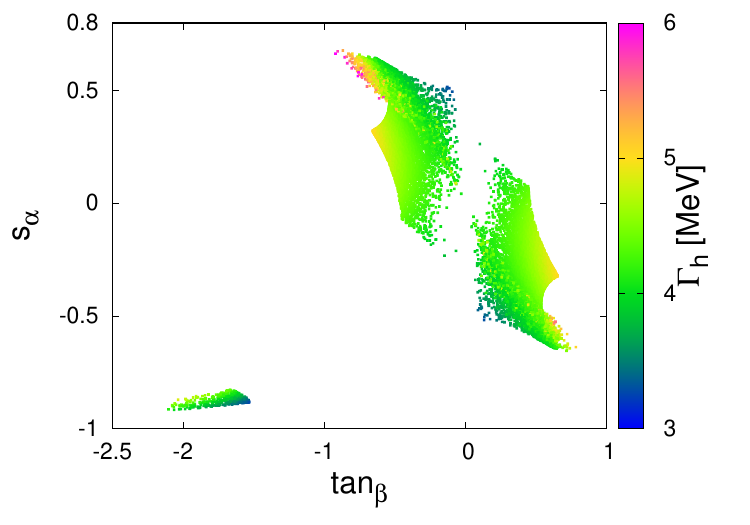}\\
 \includegraphics[width=0.49\textwidth]{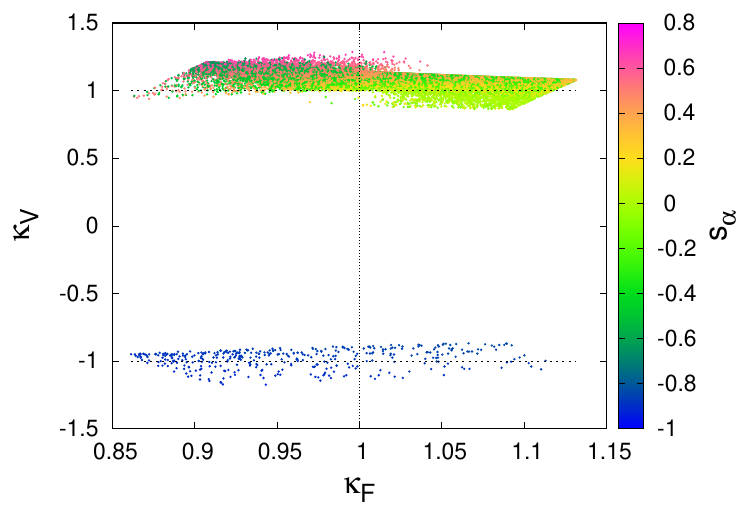}~\includegraphics[width=0.49\textwidth]{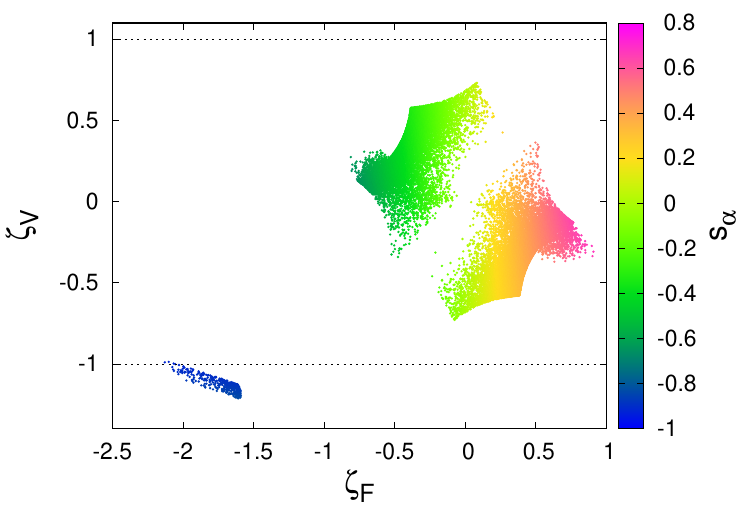}\caption{Different physical observables estimated in the GM model by taking
into account the constraints from perturbativity, vacuum stability,
electroweak precision tests, the di-photon and undetermined Higgs
decays; and the total Higgs decay width.}
\label{fig:Vac}
\end{figure}

After combining all the first step constraints, we show in Fig.~\ref{fig:Vac}
the viable parameters space and the different physical observables
using 34.7k BPs.

From Fig.~\ref{fig:Vac}, one notices a significant parameter space
comparable to the case where the SM-like Higgs is the light CP-even
eigenstate. The parameter space in the plan $\{m_3,m_5\}$ is different than the case of GM model with heavy scalar $\eta$, while in the plans $\{t_{\beta}, s_{\alpha}\}$, $\{\kappa_F, \kappa_V\}$ and $\{\zeta_F, \zeta_V\}$ they are similar. One has to notice that imposing different theoretical
and experimental constraints, especially the Higgs total width, the
Higgs signal strength modifiers and the B physics flavor constraints,
makes the parameter space well constrained. It is separated into three
distinct islands in the plan $\{t_{\beta},s_{\alpha}\}$, where the
first one corresponds to positive $t_{\beta}$ values, the second
corresponds to negative $t_{\beta}$ and positive $s_{\alpha}$ values,
while the region of negative $t_{\beta}$ and negative $s_{\alpha}$
values represents the third island. The Higgs fermions couplings ($hf\bar{f}$)
are always positive; and can be modified by a ratio up to 15\%, while
the Higgs gauge couplings ($hVV$) can be modified by few percent;
and could be negative for negative $s_{\alpha}$. The second and third
islands that were not considered previously in the literature (negative
$t_{\beta}$) correspond to negative ($hf\bar{f}$) couplings. The
magnitude of these couplings can be enhanced up to 120\%, which makes
the searches for SM-like light scalars in the light fermions channels
efficient to probe this part of the viable parameter space.

Here, one has to mention that most of the allowed light scalar mass
values are for $m_{\eta}>m_{h}/2$ due to the conflict between the
constraints from the undetermined ($h\to \eta\eta$) and di-photon
($h\to \gamma\gamma$) Higgs decays.

\begin{figure}[h]
\includegraphics[width=0.33\textwidth]{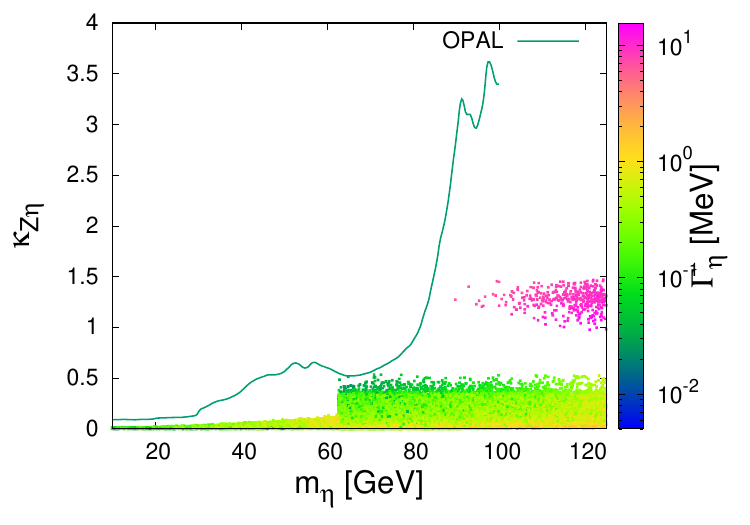}~\includegraphics[width=0.33\textwidth]{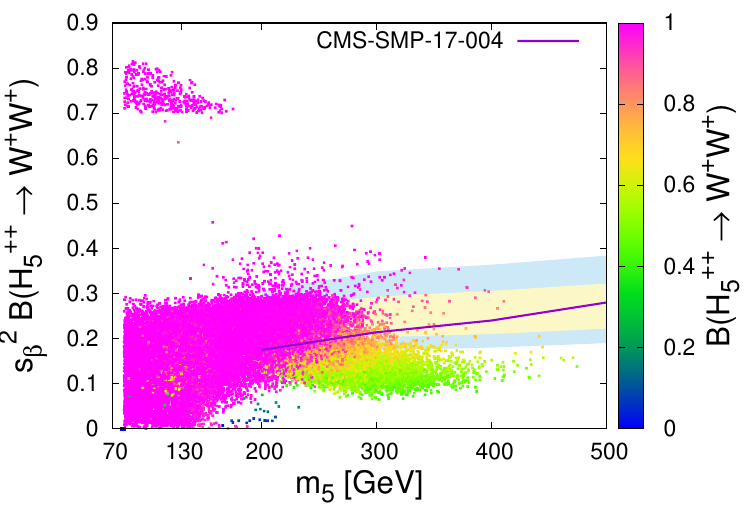}~\includegraphics[width=0.33\textwidth]{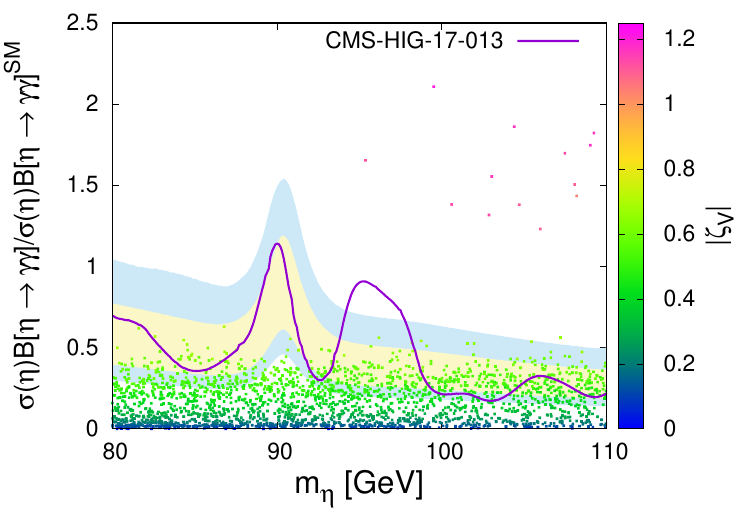}\caption{Left: the form factor (\ref{eq:kappa}) versus the light scalar mass
$m_{\eta},$where the palette shows the light scalar total decay width.
The green curve represents the OPAL bounds~\cite{OPAL:2002ifx}.
Middle: the ratio $s_{\beta}^{2}\times\mathcal{B}(H_{5}^{++}\to  W^{+}W^{+})$
compared to the CMS bounds~\cite{CMS:2017fhs}, where the yellow
(blue) region corresponds to 68\% (95\%) CL, and the palette shows
the branching ratio $\mathcal{B}(H_{5}^{++}\to  W^{+}W^{+})$.
Right: the combined cross section at 8+13 TeV $\sigma(pp\to \eta\to \gamma\gamma)$
scaled by the SM values compared to the CMS bounds for the mass range
$80\,\mathrm{GeV}<m_{\eta}<110\,\mathrm{GeV}$~\cite{CMS:2018cyk},
where the palette shows the factor (\ref{eq:kgg}) that represents
the enhancement effect on the decay $\eta\to \gamma\gamma$
due to the coupling with charged scalars. The yellow (blue) region
corresponds to 68\% (95\%) CL.}
\label{fig:Cons}
\end{figure}

Some of these 34.7k BPs are in agreement all the above mentioned constraints,
including those are considered in the second step of our analysis.
For instance, we show in Fig.~\ref{fig:Cons} some of the observables
like the form factor (\ref{eq:kappa}) that is constrained by OPAL~\cite{OPAL:2002ifx},
the ratio $s_{\beta}^{2}\times\mathcal{B}(H_{5}^{++}\to  W^{+}W^{+})$
constrained by CMS~\cite{CMS:2017fhs}; and cross section at 8+13
TeV $\sigma(pp\to \eta\to \gamma\gamma)$ constrained
by CMS~\cite{CMS:2018cyk}.

From Fig.~\ref{fig:Cons}-left and -right, one learns that the first
and second islands in Fig.~\ref{fig:Vac}-top-right is in agreement
with the OPAL bounds (i.e., the lower green island in Fig.~\ref{fig:Cons}-left
that corresponds to $|\zeta_{V}|\lesssim0.7$). The third island is
also in agreement with OPAL since it corresponds to $80~\mathrm{GeV}\leq m_{\eta}\leq120~\mathrm{GeV}$,
i.e., the pink island in Fig.~\ref{fig:Cons}-left.

From Fig.~\ref{fig:Cons}-right, one notices that the BPs with the
$\eta$ mass in the range $80~\mathrm{GeV}-110~\mathrm{GeV}$ that
are in agreement with the bound form $pp\to \eta\to \gamma\gamma$
are those that belong to the first and second islands. From Fig.~\ref{fig:Cons}-middle,
one remarks that the majority of the BPs with open decay channels $H_{5}^{++}\to  W^{+}H_{3}^{+},H_{3}^{+}H_{3}^{+}$,
which makes the branching ratio $\mathcal{B}(H_{5}^{++}\to  W^{+}W^{+})$
significantly smaller than unity. Indeed, some of the BPs with $\mathcal{B}(H_{5}^{++}\to  W^{+}W^{+})=1$
are also in agreement with the bounds on $s_{\beta}^{2}\times\mathcal{B}(H_{5}^{++}\to  W^{+}W^{+})$~\cite{CMS:2017fhs}.

\begin{figure}[h]
\includegraphics[width=0.49\textwidth]{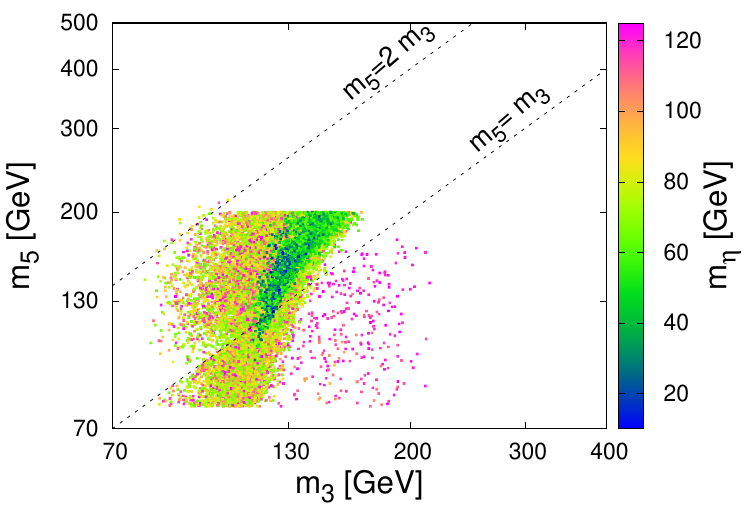}~\includegraphics[width=0.49\textwidth]{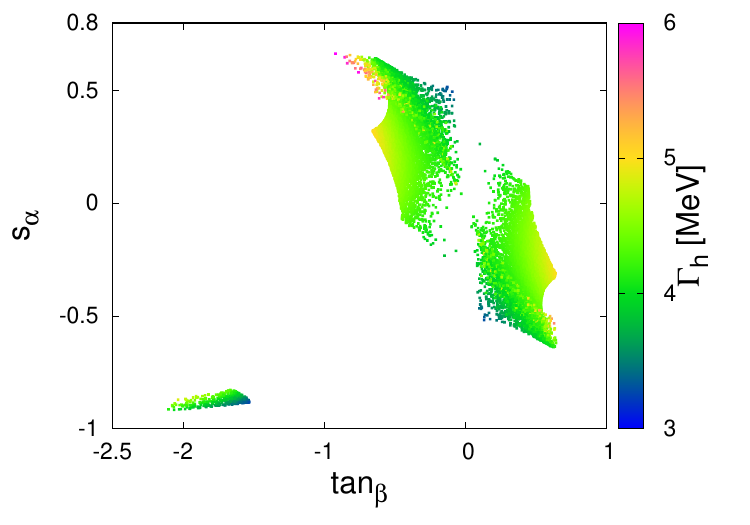}\\
 \includegraphics[width=0.49\textwidth]{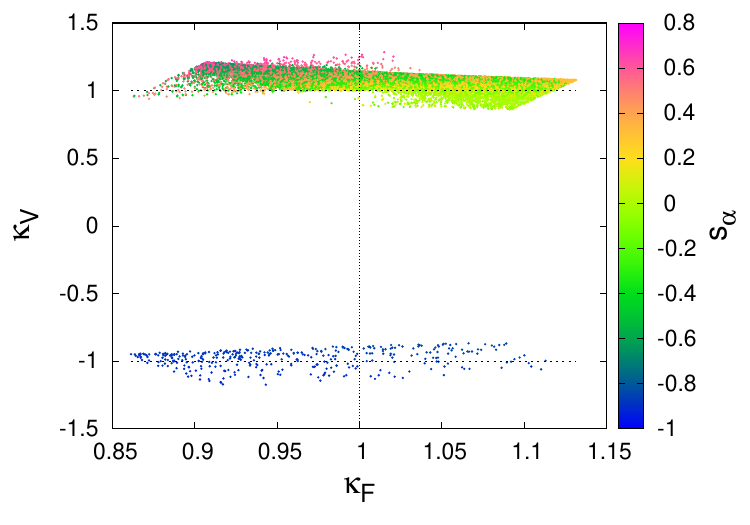}~\includegraphics[width=0.49\textwidth]{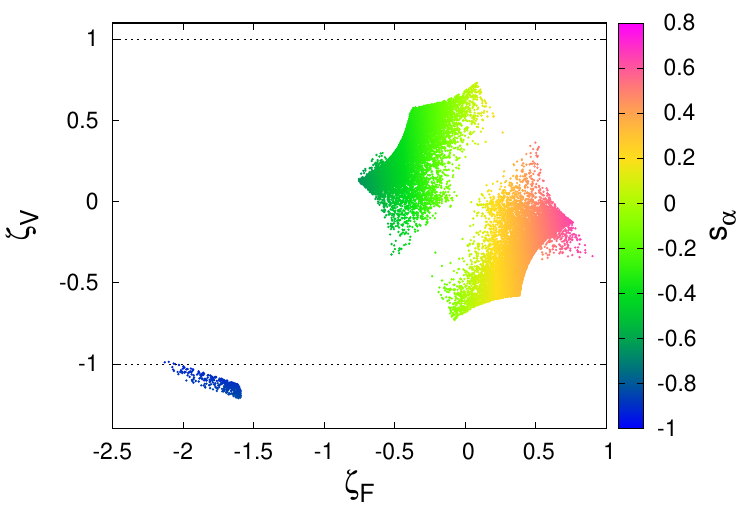}\caption{Different physical observables estimated in the GM model by taking
into account all the constraints.}
\label{fig:allowed}
\end{figure}

Clearly, the constraints we have considered in our second step analysis
seems to interesting and efficient. For instance, the constraints
from the doubly-charged Higgs bosons in the VBF channel $H_{5}^{++}\to  W^{+}W^{+}$
excludes 41.75\% of the BPs; and those from the Drell-Yan
production of a neutral Higgs boson $pp\to  H_{5}^{0}(\gamma\gamma)H_{5}^{+}$
excludes only 1.7\%. The negative direct searches of the
scalar $\eta$ exclude 0.9\% of the BPs. By combining all
these constraints, we got 44.1\% of the BPs excluded. In
Fig.~\ref{fig:allowed}, we reproduce Fig.~\ref{fig:Vac} by considering
only the viable 19.4k BPs.

From Fig.~\ref{fig:allowed}-left, one remarks that the majority
of the BPs with $m_{5}>200~\mathrm{GeV}$ are excluded, and those
that survive have the decay channels $H_{5}^{++}\to  H_{3}^{+}H_{3}^{+},H_{3}^{+}W^{+}$
open, i.e., the branching ratio for $\mathcal{B}(H_{5}^{++}\to  W^{+}W^{+})$
is significantly smaller than unity, as shown in Fig.~\ref{fig:Cons}-middle.
Most of the BPs with $m_{5}<200~\mathrm{GeV}$ are not constrained
by the searches of the same sign leptons from the doubly-charged Higgs
bosons production in the VBF channel $H_{5}^{++}\to  W^{+}W^{+}$.
Therefore, future analysis should consider the range $78~\mathrm{GeV}<m_{5}<200~\mathrm{GeV}$.
It is clear that this viable parameter space would tightened by taking
into account future analyses with more data.

In this setup, the light scalar $\eta$ can be probed due to production
modes both at the HL-LHC~\cite{HL-LHC} and at future lepton colliders
such as the ILC~\cite{Fujii:2017vwa}, CLIC~\cite{Abramowicz:2016zbo},
the CEPC~\cite{Chen:2016zpw} and the FCCee~\cite{Dawson:2013bba}
running at the Z-pole and above the Zh production threshold. In future
searches, more severe constraints on the parameters $m_{\eta},~\kappa_{V,F},~\zeta_{V,F}$
would be established. Distinguish the GM model among other similar
models that involves a light scalar such by just relying on the Higgs
measurements and the $\eta$-discovery (or $\eta$ negative searches)
is not trivial. However, in the GM model, a correlation between the
scalar-gauge and scalar-fermions couplings is absent unlike most of
the SM multi-scalar extensions, i.e., no correlation between $\kappa_{V}-\kappa_{F}$
and $\zeta_{V}-\zeta_{F}$ exists. This feature could help to probe
the GM model in some cases where both ratios $\kappa_{V}/\kappa_{F}$
and $\zeta_{V}/\zeta_{F}$ are significantly different than unity.
In addition, the custodial triplet and fiveplet members have many
production modes, which makes the GM model more interesting for any
search.

\section{Conclusion\label{sec:Conclusion}}

In this work, we studied the GM scalar sector in the case where the
SM-like Higgs corresponds to the heavy CP-even eigenstate. We have
shown the viability of an important region of the parameter space,
that is significant as the case of the light CP-even scalar to be
the SM-like Higgs. In our analysis, we considered the constraints
from perturbativity, unitarity, boundness from below, the electroweak
precision tests, the di-photon and undetermined Higgs decays; and
the total Higgs decay width. For this we generated around 34.7k BPs
that fulfill all the previously mentioned constraints. In addition,
we have imposed more bounds from the searches for (1) doubly-charged
Higgs bosons in the VBF channel $H_{5}^{++}\to  W^{+}W^{+}$,
(2) Drell-Yan production of a neutral Higgs boson $pp\to  H_{5}^{0}(\gamma\gamma)H_{5}^{+}$,
and for the light scalars by ATLAS and CMS in different final states
such as (3) $pp\to  h\to \eta\eta\to 4\gamma,2\mu2\tau,2\mu2b,2\tau2b$.
We found that only 55.9\% of the BPs survives against these
three constraints, where they exclude 41.75\%, 1.7\%,
and 0.9\% of the BPs, respectively.

After imposing all constraints, we found that the negative searches
of the doubly-charged Higgs bosons in the VBF channel $H_{5}^{++}\to  W^{+}W^{+}$
exclude all BPs with $\mathcal{B}(H_{5}^{++}\to  W^{+}W^{+})=1$
and $m_{5}>200\,\mathrm{GeV}$, while all surviving BPs have the decay
channels $H_{5}^{++}\to  H_{3}^{+}H_{3}^{+},H_{3}^{+}W^{+}$
open, $m_{5}>2m_{3}$. It is important if the future searches for
doubly charged scalars would consider masses below $200\,\mathrm{GeV}$
to probe this scenario. In the near future, the coming analyses with
more data will make the parameter space more constrained.

The light scalar $\eta$ can be produced and detected at future hadron/lepton
colliders, which is the case of many light scalars in multi-scalar
models. Then, more severe constraints on the parameters $m_{\eta},~\kappa_{V,F},~\zeta_{V,F}$
would be established. However, distinguishing the GM light scalar
among other scalars involved in other multi-scalar models could be
possible; and relies on the measurement of the $\eta$ couplings to
gauge bosons ($\zeta_{V}$) and fermions ($\zeta_{F}$). The reason
for this is that, unlike in many multi-scalar models, a correlation
between the scalar-gauge and scalar-fermions couplings is absent in
the GM model. Although, considering some future neutral/charged scalar
searches may put more constraints on the custodial triplet and fiveplet
members, which makes the GM model more interesting.

\textbf{Acknowledgments}: This work is funded by the University of
Sharjah under the research projects No 21021430100 ``\textit{Extended
Higgs Sectors at Colliders: Constraints \& Predictions}'' and No
21021430107 ``\textit{Hunting for New Physics at Colliders}''.

\end{document}